 \documentclass[final,5p,times,twocolumn]{elsarticle} 

\usepackage{amsfonts}
\usepackage{amsmath}
\usepackage{amssymb}
\usepackage{booktabs}
\usepackage{caption}
\usepackage{color}
\usepackage{comment}
\usepackage{graphicx}
\usepackage{hyperref}
\usepackage[utf8]{inputenc} 
\usepackage{subfig}
\usepackage{xspace}

\newcommand{\pygbe}{\texttt{PyGBe}\xspace}
\newcommand{\gb}{{\small G\,B1\,D4$^\prime$}\xspace}
\newcommand{\gmres}{\textsc{gmres}\xspace}
\newcommand{\bem}{\textsc{bem}\xspace}
\newcommand{\ses}{\textsc{ses}\xspace}
\newcommand{\sam}{\textsc{sam}}
\newcommand{\gpu}{\textsc{gpu}}
\newcommand{\cpu}{\textsc{cpu}}
\newcommand{\apbs}{\textsc{apbs}\xspace}
\newcommand{\nvidia}{\textsc{nvidia}\xspace}
\newcommand{\msms}{\texttt{\textsc{msms}}\xspace}
\newcommand{\amber}{\texttt{\textsc{amber}}\xspace}
\newcommand{\ccby}{\textsc{cc-by}\xspace}

\graphicspath{{figs/}} 

\journal{Computer Physics Communications}

\begin{document}

\begin{frontmatter}



\title{Poisson-Boltzmann model for protein-surface electrostatic interactions and grid-convergence study using the \pygbe code}

\author[bu,usm]{Christopher D. Cooper\corref{cdc}}
\ead{christopher.cooper@usm.cl}

\author[gwu]{Lorena A.~Barba\corref{lab}}
\ead{labarba@gwu.edu}

\address[bu]{Department of Mechanical Engineering, Boston University, Boston, MA.}
\address[usm]{Department of Mechanical Engineering, Universidad T\'ecnica Federico Santa Mar\'ia, Valpara\'iso, Chile.}
\address[gwu]{Department of Mechanical \& Aerospace Engineering, The George Washington University, Washington, D.C.}
\cortext[lab]{\href{mailto:labarba@gwu.edu}{labarba@gwu.edu}}


\begin{abstract}

Interactions between surfaces and proteins occur in many vital processes and are crucial in biotechnology: the ability to control specific interactions is essential in fields like biomaterials, biomedical implants and biosensors. In the latter case, biosensor sensitivity hinges on ligand proteins adsorbing on bioactive surfaces with a favorable orientation, exposing reaction sites to target molecules.
Protein adsorption, being a free-energy-driven process, is difficult to study experimentally. This paper develops and evaluates a computational model to study electrostatic interactions of proteins and charged nanosurfaces, via the Poisson-Boltzmann equation.
We extended the implicit-solvent model used in the open-source code PyGBe to include surfaces of imposed charge or potential. This code solves the boundary integral formulation of the Poisson-Boltzmann equation, discretized with surface elements. PyGBe has at its core a treecode-accelerated Krylov iterative solver, resulting in $O(N \log N)$ scaling, with further acceleration on hardware via multi-threaded execution on \gpu s. It computes solvation and surface free energies, providing a framework for studying the effect of electrostatics on adsorption.
We then derived an analytical solution for a spherical charged surface interacting with a spherical molecule, then completed a grid-convergence study to build evidence on the correctness of our approach. The study showed the error decaying with the average area of the boundary elements, i.e., the method is $O(1/N)$, which is consistent with our previous verification studies using PyGBe.
We also studied grid-convergence using a real molecular geometry (protein \gb), in this case using Richardson extrapolation (in the absence of an analytical solution) and confirmed the $O(1/N)$ scaling in this case.
PyGBe is open-source under an MIT license and is hosted under version control at \href{https://github.com/barbagroup/pygbe}{https://github.com/barbagroup/pygbe}. In addition, we prepared ``reproducibility packages''  to supplement this paper, consisting of running and post-processing scripts in Python to allow replication of the grid-convergence studies, all the way to generating the final plots, with a single command.
\end{abstract}


\end{frontmatter}


\section{Introduction}\label{sec:intro}

Proteins interacting with solid surfaces fundamentally appear in many biological processes. Adsorption serves a key function in natural activities, like blood coagulation, and  in biotechnologies like tissue engineering, biomedical implants and biosensors.
A full understanding of protein-surface interactions has remained elusive \cite{Gray2004,RabeVerdesSeegel2011}, but adsorption mechanisms are governed by surface energy and often the dominant effect is electrostatics. As a free-energy-driven process, protein-surface interaction is difficult to study experimentally \cite{MijajlovicETal2013}, and thus simulations offer a good alternative. Full atomistic molecular dynamics simulations demand large amounts of computing effort, however, so we often must resort to other methods.

Protein electrostatics can be studied via modeling approaches using the Poisson-Boltzmann equation and implicit-solvent representations. These models  are popular for computing solvation energies in protein systems \cite{RouxSimonson1999,Bardhan2012}, but few studies have included the effect of surfaces. Lenhoff and co-workers studied surface-protein interactions using continuum models discretized with boundary-element \cite{YoonLenhoff1992,RothLenhoff1993,AsthagiriLenhoff1997} and finite-difference methods \cite{YaoLenhoff2004,YaoLenhoff2005}, in the context of ion-exchange chromatography. They realized that van der Waals effects can be neglected for realistic molecular geometries \cite{RothNealLenhoff1996} and that the model is adequate as long as conformational changes in the protein are slight \cite{YaoLenhoff2004,YaoLenhoff2005}. 

The aim of this work is to develop and assess a computational model to simulate proteins near engineered surfaces of fixed charge, using implicit-solvent electrostatics.
We have added the capability of modeling a protein near a charged surface to our code \pygbe, an open-source code\footnote{\url{https://github.com/barbagroup/pygbe}}  that solves the Poisson-Boltzmann equations via an integral formulation, using a fast multipole algorithm and \gpu\ hardware acceleration.  Previously, we verified and validated \pygbe in its use to obtain solvation and binding energies, by comparing with analytical solutions of the equations and with results obtained using the well-known \apbs software \cite{CooperBarba-share154331,CooperBardhanBarba2013}. 
In the present work, we derived an analytical solution for a spherical molecule interacting with a spherical surface of prescribed charge, and used it to verify the code in its new application and study numerical convergence.
Using the newly extended code, we also studied the interaction between protein \gb and a solid surface of imposed charge, 
and conduct a grid-convergence study using this more realistic surface geometry.

We intend our new modeling tool to be useful in studying the behavior of proteins as they adsorb on surfaces that have been functionalized with  self-assembled monolayers (\sam), which are modeled within an implicit-solvent framework as surfaces with prescribed charge. 
One application is biosensing, where the target molecules are captured on the sensor via ligand molecules (for which antibodies are a common choice). Favorable orientations of ligand molecules lead to greatly enhanced sensitivity of biosensors \cite{TajimaTakaiIshihara2011,TrillingBeekwilderZuilhof2013}, because binding sites need to be physically accessible to the targets. Studies of protein orientation near charged surfaces might look at how orientation can be influenced by engineering decisions regarding surface preparation, to aid the design of better biosensors. We explore this application in a companion publication that obtains probability of orientations for an antibody near a surface, as function of changing conditions on charge and ionic strength \cite{CooperBarba2015b}. 
In this paper, we present the details of a new analytical solution for spherical charged surfaces and molecules, grid-convergence studies for the interaction free energy in this case, and grid-convergence studies for protein \gb alone and interacting with a charged surface. The detailed analysis of the model is complemented with a diligent effort for reproducibility and we deposit both input and results data in accessible and permanent archival storage, in addition to the open-source code.

\section{Implicit-solvent model for proteins near charged surfaces} \label{sec:implicit_solvent}

\begin{figure}
   \includegraphics[width=0.49\textwidth]{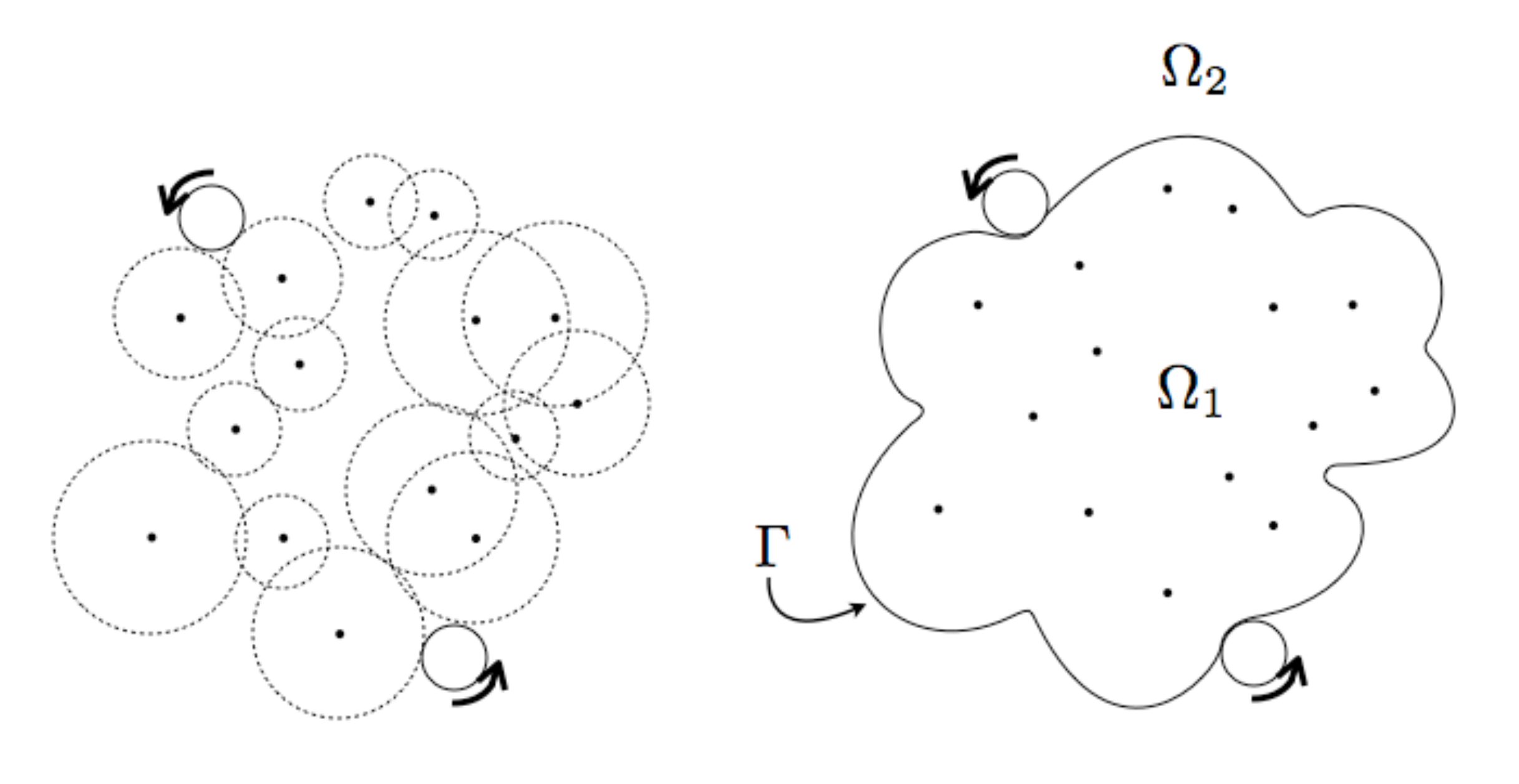} 
   \caption{Sketch of the process for generating a solvent-excluded surface (\ses): a protein molecule contains a set of atoms that define a radius upon applying a force field and a probe the size of a water molecule is rolled to define the  \ses. $\Omega_1$ is the protein region and $\Omega_2$ the solvent region.}
   \label{fig:forcefield-ses}
\end{figure}

The implicit-solvent model uses continuum electrostatics to describe the mean-field potential in a molecular system. A typical system consists of a protein in a solvent, defining two regions: inside and outside the protein, with an interface marked by the solvent-excluded surface (\ses).  The \ses, beyond which a water molecule cannot penetrate into the protein, can be generated by rolling a (virtual) spherical probe of the size of a water molecule around the protein (see Figure \ref{fig:forcefield-ses}). Inside the protein, the domain has low permittivity ($\epsilon= 2\text{ to }4$) and there are point charges located at the positions of the atoms. The solvent region, representing water with salt, has a permittivity of $\epsilon \approx 80$. A system of partial differential equations models this situation, with a Poisson equation governing inside the protein and a linearized Poisson-Boltzmann equation governing in the solvent region. Appropriate interface conditions on the \ses express the continuity of the potential and electric displacement, completing the mathematical formulation.

\begin{figure}
   \includegraphics[width=0.45\textwidth]{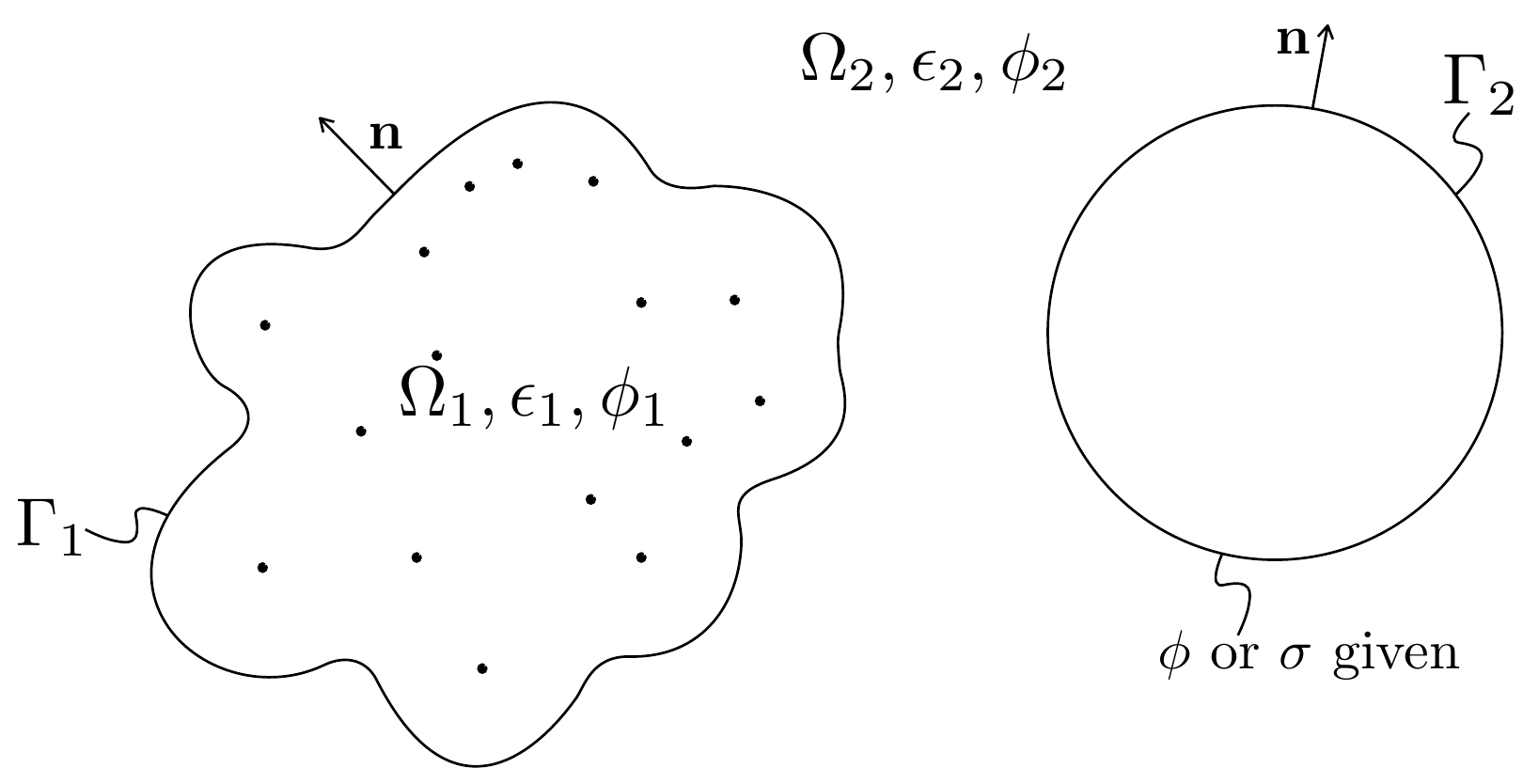} 
   \caption{Sketch of a molecule interacting with a surface: $\Omega_1$ is the protein, $\Omega_2$ the solvent region, $\Gamma_1$ is the  \ses and $\Gamma_2$ a nanosurface with imposed charge or potential.}
   \label{fig:molecule_surface}
\end{figure}

This model has been widely applied to investigate interactions between molecules, such as in protein-ligand binding. We are interested here in an extension of the model to consider interactions between  proteins and surfaces with an imposed potential or charge. This new setup is sketched in Figure \ref{fig:molecule_surface}, and is described mathematically by the following equations:

\begin{align} \label{eq:pde}
\nabla^2 \phi_1(\mathbf{r}) &= - \sum_k \frac{q_k}{\epsilon_1} \delta(\mathbf{r},\mathbf{r}_k) \ \text{ in solute $(\Omega_1)$,}  \nonumber \\ 
\nabla^2\phi_2 (\mathbf{r}) &= \kappa^2 \phi_2(\mathbf{r}) \quad \qquad \ \ \text{ in solvent $(\Omega_2)$,}  \nonumber \\ 
\phi_1 &=\phi_2 \qquad \qquad \qquad \text{ on interface $\Gamma_1$,}  \nonumber \\ 
\epsilon_1 \frac{\partial \phi_1}{\partial \mathbf{n}} &= \epsilon_2 \frac{\partial \phi_2}{\partial \mathbf{n}} \nonumber \\
\phi_2 = \phi_0 &\text{ or } -\epsilon_2 \frac{\partial \phi_2}{\partial \mathbf{n}} = \sigma_0 \ \ \text{ on surface $\Gamma_2$,} 
\end{align}

\noindent Here, $\phi_i$ is the potential corresponding to the region $\Omega_i$ with permittivity $\epsilon_i$, and $\phi_0$ and $\sigma_0$ are the set potential or charge on the nanosurface. 

\paragraph*{Boundary integral formulation} \label{sec:bie}

We express the system of partial-differential equations in  \eqref{eq:pde} by the corresponding integral equations along the interface and the nanosurface, $\Gamma_1$ and $\Gamma_2$. Many authors have used the boundary-integral representation of the implicit-solvent model to compute solvation energies of proteins \cite{YoonLenhoff1990, Juffer1991a, LuETal2006, BajajETal2011, AltmanBardhanWhiteTidor09, GengKrasny2013, CooperBardhanBarba2013}, but apart from work led by Lenhoff \cite{YoonLenhoff1992}, we know of no studies that account for interacting surfaces in the system. 

Consider the setting in Figure \ref{fig:molecule_surface} with prescribed potential at $\Gamma_2$. The application of Green's second identity on the first two equations of \eqref{eq:pde} yields:
\begin{align} \label{eq:green_identity}
\phi_{1}+ K_{L}^{\Omega_1}(\phi_{1,\Gamma_1}) -  V_{L}^{\Omega_1} \left(\frac{\partial}{\partial \mathbf{n}}  \phi_{1,\Gamma_1}  \right) &  = \nonumber\\
 \frac{1}{\epsilon_1} \sum_{k=0}^{N_q}  \frac{q_k}{4\pi|\mathbf{r}_{\Omega_1} - \mathbf{r}_k|} &  \quad \text{on $\Omega_1$,} \nonumber \\ \nonumber \\
\phi_{2} - K_{Y}^{\Omega_2}(\phi_{2,\Gamma_1}) + V_{Y}^{\Omega_2} \left( \frac{\partial}{\partial \mathbf{n}} \phi_{2,\Gamma_1} \right) - & K_{Y}^{\Omega_2}(\phi_{2,\Gamma_2})  \nonumber \\
 + V_{Y}^{\Omega_2}  \left( \frac{\partial}{\partial \mathbf{n}} \phi_{2,\Gamma_2} \right) & = 0 \quad \text{on $\Omega_2$,}
\end{align}

\noindent where $\phi_{i,\Gamma_j} = \phi_i(\mathbf{r}_{\Gamma_j})$ is the potential in region $\Omega_i$ evaluated at the surface $\Gamma_j$. $K$ and $V$ are defined as

\begin{align} \label{eq:layers}
K_{L/Y}^{\Omega_i}(\phi_{i,\Gamma_j}) &= \oint_{\Gamma_j} \frac{\partial}{\partial \mathbf{n}} \left[ G_{L/Y}(\mathbf{r}_{\Omega_i},\mathbf{r}_{\Gamma_j}) \right]\phi_{i,\Gamma_j} \, \mathrm{d} \Gamma, \nonumber \\
V_{L/Y}^{\Omega_i} \left( \frac{\partial}{\partial \mathbf{n}} \phi_{i,\Gamma_j} \right) &= \oint_{\Gamma_j} \frac{\partial}{\partial \mathbf{n}} \phi_{i,\Gamma_j} G_{L/Y}(\mathbf{r}_{\Omega_i},\mathbf{r}_{\Gamma_j})  \, \mathrm{d} \Gamma,
\end{align}

\noindent corresponding to the double- and single-layer potentials of $\phi_{i,\Gamma_j}$ and $\frac{\partial}{\partial \mathbf{n}} \phi_{i,\Gamma_j}$ evaluated in the region $\Omega_i$. The functions $G_L$ and $G_Y$ are the free-space Green's functions of the Poisson (Laplace kernel) and linearized Poisson-Boltzmann (Yukawa kernel) equations, respectively:

\begin{align} \label{eq:free-space}
G_L(\mathbf{r}_{\Omega_1},\mathbf{r}_{\Gamma_1}) &= \frac{1}{4\pi|\mathbf{r}_{\Omega_1} - \mathbf{r}_{\Gamma_1}|}, \nonumber \\
G_Y(\mathbf{r}_{\Omega_2},\mathbf{r}_{\Gamma_2}) &= \frac{\exp \left( -\kappa |\mathbf{r}_{\Omega_1} - \mathbf{r}_{\Gamma_1}|\right)}{4\pi|\mathbf{r}_{\Omega_1} - \mathbf{r}_{\Gamma_1}|}.
\end{align}

\noindent We then take the limits 
$\mathbf{r}_{\Omega_1}\!\to\!\mathbf{r}_{\Gamma_1}$, 
$\mathbf{r}_{\Omega_2}\!\to\!\mathbf{r}_{\Gamma_1}$, 
$\mathbf{r}_{\Omega_2}\!\to\!\mathbf{r}_{\Gamma_2}$ on Equation \eqref{eq:green_identity}, and apply the boundary conditions:
$\phi_{1,\Gamma_1} = \phi_{2,\Gamma_1}$, 
$\epsilon_1\frac{\partial}{\partial \mathbf{n}} \phi_{1,\Gamma_1} =  \epsilon_2\frac{\partial}{\partial \mathbf{n}} \phi_{2,\Gamma_1} $ 
and $\phi_{2,\Gamma_2} = \phi_0$ to get the following system of boundary equations:

\vspace{1em}

%
\begin{align} \label{eq:integral_eq}
\frac{\phi_{1,\Gamma_1}}{2}+ K_{L}^{\Gamma_1}(\phi_{1,\Gamma_1}) - &V_{L}^{\Gamma_1} \left(\frac{\partial}{\partial \mathbf{n}}\phi_{1,\Gamma_1} \right)\nonumber\\  
&= \frac{1}{\epsilon_1} \sum_{k=0}^{N_q} \frac{q_k}{4\pi|\mathbf{r}_{\Gamma_1} - \mathbf{r}_k|}  \quad \text{on $\Gamma_1$,} \nonumber \\ 
\frac{\phi_{1,\Gamma_1}}{2} - K_{Y}^{\Gamma_1}(\phi_{1,\Gamma_1}) +  &\frac{\epsilon_1}{\epsilon_2} V_{Y}^{\Gamma_1} \left( \frac{\partial}{\partial \mathbf{n}} \phi_{1,\Gamma_1} \right)\nonumber \\ 
- K_{Y}^{\Gamma_1}&(\phi_{0})  + V_{Y}^{\Gamma_1} \left( \frac{\partial}{\partial \mathbf{n}} \phi_{2,\Gamma_2} \right)  = 0 \quad \text{on $\Gamma_1$,} \nonumber \\ 
- K_{Y}^{\Gamma_2}(\phi_{1,\Gamma_1}) + \frac{\epsilon_1}{\epsilon_2} V_{Y}^{\Gamma_2} &\left( \frac{\partial}{\partial \mathbf{n}} \phi_{1,\Gamma_1} \right)\nonumber \\ 
+ \frac{\phi_{0}}{2} - &K_{Y}^{\Gamma_2}(\phi_{0}) +  V_{Y}^{\Gamma_2} \left( \frac{\partial}{\partial \mathbf{n}} \phi_{2,\Gamma_2} \right)  = 0 \quad \text{on $\Gamma_2$.}
\end{align}

\noindent Rearranging terms, we write Equation \eqref{eq:integral_eq} in matrix form, as follows:
 \begin{align} \label{eq:matrix_phi}
 \left[
    \begin{matrix} 
       \frac{1}{2} + K_{L}^{\Gamma_1} & -V_{L}^{\Gamma_1} & 0 \vspace{0.2cm}\\
       \frac{1}{2} - K_{Y}^{\Gamma_1} &  \frac{\epsilon_1}{\epsilon_2} V_{Y}^{\Gamma_1} & V_{Y}^{\Gamma_1} \vspace{0.2cm} \\
       - K_{Y}^{\Gamma_2} & \frac{\epsilon_1}{\epsilon_2} V_{Y}^{\Gamma_2} & V_{Y}^{\Gamma_2} \\
    \end{matrix}
    \right] \left[ 
    \begin{matrix} 
       \phi_{1,\Gamma_1} \vspace{0.2cm} \\
       \frac{\partial}{\partial \mathbf{n}} \phi_{1,\Gamma_1} \vspace{0.2cm}\\
       \frac{\partial}{\partial \mathbf{n}} \phi_{2,\Gamma_2}\\
    \end{matrix} 
     \right] =  & \nonumber \\
    \left[
    \begin{matrix} 
       \sum_{k=0}^{N_q} \frac{q_k}{4\pi|\mathbf{r}_{\Gamma_1} - \mathbf{r}_k|} \vspace{0.2cm} \\
        K_{Y}^{\Gamma_1}(\phi_{0}) \vspace{0.2cm} \\
        -\left(\frac{1}{2} - K_{Y}^{\Gamma_2}\right) (\phi_0)
    \end{matrix}
    \right].
 \end{align}

\medskip
\noindent If the surface $\Gamma_2$ has prescribed charge, corresponding to a Neumann boundary condition, $-\epsilon_2\frac{\partial}{\partial \mathbf{n}} \phi_{2,\Gamma_2} = \sigma_0$, the equivalent derivation yields

 \begin{align} \label{eq:matrix_dphi}
 \left[
    \begin{matrix} 
       \frac{1}{2} + K_{L}^{\Gamma_1} & -V_{L}^{\Gamma_1} & 0 \vspace{0.2cm}\\
       \frac{1}{2} - K_{Y}^{\Gamma_1} &  \frac{\epsilon_1}{\epsilon_2} V_{Y}^{\Gamma_1} & -K_{Y}^{\Gamma_1} \vspace{0.2cm} \\
       - K_{Y}^{\Gamma_2} & \frac{\epsilon_1}{\epsilon_2} V_{Y}^{\Gamma_2} & \left(\frac{1}{2} - K_{Y}^{\Gamma_2}\right) \\
    \end{matrix}
    \right] \left[ 
    \begin{matrix} 
       \phi_{1,\Gamma_1} \vspace{0.2cm} \\
       \frac{\partial}{\partial \mathbf{n}} \phi_{1,\Gamma_1} \vspace{0.2cm}\\
       \phi_{2,\Gamma_2}\\
    \end{matrix} 
     \right] =   \nonumber \\
    \left[
    \begin{matrix} 
       \sum_{k=0}^{N_q} \frac{q_k}{4\pi|\mathbf{r}_{\Gamma_1} - \mathbf{r}_k|} \vspace{0.2cm} \\
        -V_{Y}^{\Gamma_1} \left( \frac{\partial}{\partial \mathbf{n}} \phi_{2,\Gamma_2} \right) \vspace{0.2cm} \\
        -V_{Y}^{\Gamma_2} \left( \frac{\partial}{\partial \mathbf{n}} \phi_{2,\Gamma_2} \right)
    \end{matrix}
    \right].
 \end{align}
 

The formulation detailed in this section differs from the work by Lenhoff and co-workers \cite{YoonLenhoff1992,RothLenhoff1993} because they consider an infinite charged surface, modeled using a modified Green's function to account for the half-space domain. Lenhoff's approach has the advantage that the charged surface does not require a mesh, but cannot be applied to any situation where the surface has non-planar geometry. An infinite surface may not be a good model if the size of a device is comparable to the protein's, like it happens with nano-structures.
In that case, one needs to be able to represent the detailed geometry of a device via a surface mesh, and satisfy boundary conditions there, as detailed above.

The system in \eqref{eq:matrix_dphi} can be extended to account for Stern layers and solvent-filled cavities by adding more surfaces or interfaces. In our code, we deal with multiple surfaces in the manner presented by Altman and co-workers \cite{AltmanBardhanWhiteTidor09}, as described in our previous paper \cite{CooperBardhanBarba2013}.

\section{Methods}\label{sec:methods}


\subsection{Discretization}

To numerically solve the system in \eqref{eq:matrix_dphi}, we discretize the boundaries into flat triangular panels and assume that $\phi$ and $\frac{\partial \phi}{\partial \mathbf{n}}$ are constant within those panels. The discretized form of the integral operators is as follows:
\begin{align} \label{eq:layers_disc}
&K_{L,\text{disc}}^{\mathbf{r}_i}\left(\phi(\mathbf{r}_{\Gamma})\right) =  \sum_{j=1}^{N_p}\phi(\mathbf{r}_{\Gamma_j})\int_{\Gamma_j} \frac{\partial}{\partial \mathbf{n}} \left[ G_L(\mathbf{r}_{i},\mathbf{r}_{\Gamma_j}) \right]\mathrm{d} \Gamma_j,  \nonumber \\
&V_{L,\text{disc}}^{\mathbf{r}_i} \left( \frac{\partial}{\partial \mathbf{n}} \phi(\mathbf{r}_{\Gamma}) \right) = \sum_{j=1}^{N_p} \frac{\partial}{\partial \mathbf{n}} \phi(\mathbf{r}_{\Gamma_j}) \int_{\Gamma_j} G_L(\mathbf{r}_{i},\mathbf{r}_{\Gamma_j})  \mathrm{d} \Gamma_j,
\end{align}

\noindent where $N_p$ is the number of discretization elements on $\Gamma$, and $\phi(\mathbf{r}_{\Gamma_j})$ and $\frac{\partial}{\partial \mathbf{n}} \phi(\mathbf{r}_{\Gamma_j})$ are the constant values of $\phi$ and $\frac{\partial \phi}{\partial \mathbf{n}}$ on panel $\Gamma_j$ (we are somewhat abusing the nomenclature here by reusing the symbol $\Gamma$, which previously referred to the complete surface). By collocating $\mathbf{r}_i$ on the center of each panel, we get a linear system of equations that look just like \eqref{eq:matrix_phi} or \eqref{eq:matrix_dphi}, but the coefficient matrix is formed by sub-matrices of size $N_p \times N_p$ rather than integral operators. Each element of a sub-matrix is an integral over one panel $\Gamma_j$, with $\mathbf{r}_i$ located at the center of the collocation panel $\Gamma_i$, as follows:

\begin{align} \label{eq:layers_element}
K_{L,ij} &= \int_{\Gamma_j} \frac{\partial}{\partial \mathbf{n}} \left[ G_L(\mathbf{r}_{\Gamma_i},\mathbf{r}_{\Gamma_j}) \right]\mathrm{d} \Gamma_j, \nonumber \\
V_{L,ij} &= \int_{\Gamma_j} G_L(\mathbf{r}_{\Gamma_i},\mathbf{r}_{\Gamma_j})  \mathrm{d} \Gamma_j.
\end{align}

The terms on the right-hand side and the unknown vectors in the discretized form of Equation \eqref{eq:matrix_phi} are sub-vectors of size $N_p$. In this case, each element is the evaluation on the collocation panel $\Gamma_i$, written as
\begin{align} \label{eq:vector_disc}
\phi_{1,\Gamma_1} &= \phi_1(\mathbf{r}_i), \nonumber \\
\frac{\partial}{\partial \mathbf{n}}\phi_{1,\Gamma_1} &= \frac{\partial}{\partial \mathbf{n}}\phi_1(\mathbf{r}_i), \nonumber \\
\sum_{k=0}^{N_q} \frac{q}{4\pi|\mathbf{r}_{\Gamma_1} - \mathbf{r}_k|} &= \sum_{k=0}^{N_q} \frac{q}{4\pi|\mathbf{r}_i - \mathbf{r}_k|},
\end{align}

\noindent where $\mathbf{r}_i$ is located at the center of panel $\Gamma_i$.

In our numerical solution, integrals are calculated in three possible ways, depending on how close the panel is to the collocation point. When the collocation point is inside the element being integrated, we use a semi-analytical technique, with Gauss points placed along the edges of the element \cite[p.~49, ff.]{HessSmith1967}\cite{ZhuHuangSongWhite2001}. If the integrated element is closer than $2L$ from the collocation point ---where $L = \sqrt{2\cdot A_j}$ for $A_j$ the area of panel $j$--- we use a fine Gauss quadrature rule, with 19 or more points per element. Beyond a distance of $2L$, elements have only 1, 3, 4 or 7 Gauss points, depending on the case.

\subsection{Treecode-accelerated boundary element method}

Most modern implementations of the boundary element method (\bem) use Krylov methods to solve the linear system, usually a general minimal residual method (\gmres), which is agnostic to the structure of the matrix. In practice, Krylov solvers for \bem require $O(n \cdot N_p^2)$ operations to obtain the unknown vector, where $n$ is the number of iterations to get a desired residual, and is much smaller than $N_p$. The $O(N^2)$ scaling is given by a matrix-vector product (with a dense matrix) done in every iteration; this is the most time-consuming part of the algorithm, and makes \bem prohibitive for more than a few thousand discretization elements. 

But when we inspect the approximation of the integrals in  \eqref{eq:layers_element} with Gauss quadrature rules, we see that the matrix-vector product has the form of an $N$-body problem, similar to gravitational potential calculations in planetary systems. In this case, the Gauss quadrature points act analogously to planets (sources of mass) and the collocation points are analogous to the locations where the gravitational potential is computed (targets points). There are several ways to accelerate this kind of computations, for example fast-multipole methods \cite{GreengardRokhlin1987}, treecodes \cite{BarnesHut1986}, and fast-Fourier-transform methods \cite{PhillipsWhite1997}.
In our numerical solution (developed as the open-source code \pygbe), we accelerate the $N$-body calculation with a treecode \cite{BarnesHut1986,LiJohnstonKrasny2009}, making this part of the algorithm scale as $O(N\log N)$ rather than $O(N^2)$. 

The treecode algorithm groups the sources and targets in a tree-structured set of boxes and approximates interactions between far-away boxes using a series expansion---a Taylor series, in our case. This allows for controllable accuracy that depends on the number of terms used in the expansion and the multipole-acceptance criterion that defines the threshold where the distance between source and target is far enough to approximate the interactions with expansions. Details of our implementation of the treecode in \pygbe can be found in our previous work \cite{CooperBarba-share154331}.

\subsection{Energy calculation} \label{sec:energy}

Figure \ref{fig:molecule_surface} shows a system with three types of free energy: Coulombic energy from the point charges, surface energy due to $\Gamma_2$ and solvation energy. The Coulombic energy arises simply from the Coulomb interactions of all point charges. This section describes how we compute the other two components of free energy in the boundary-element framework.

\medskip

\paragraph*{Solvation free energy}

When a protein is in a solvated state, surrounded by water molecules that have become polarized, its free energy differs from its state \emph{in vacuo} by an amount known as the solvation energy. Its free energy again differs in the presence of other structures in the solvent, e.g., other proteins or charged surfaces. In this work we use the term solvation energy to more broadly mean the change in free energy of the protein from its state in a vacuum, to its state in the solvent with any other components or structures. In single-molecule settings, this definition of solvation energy coincides with the energy required to solvate the molecule. 

To calculate the solvation energy, the total minus the Coulomb potential is applied inside the protein, i.e.,

\begin{align} \label{eq:solv_energy}
F_{\text{solv}} &= \frac{1}{2} \int_{\Omega} \rho \,(\phi_{\text{total}} - \phi_{\text{Coulomb}}) \\
&= \sum_{k=0}^{N_q} q_k (\phi_{\text{total}} - \phi_{\text{Coulomb}})(\mathbf{r}_k),
\end{align}

\noindent where $\rho$ is the charge distribution, consisting of point charges (which transforms the integral into a sum). 
The total minus Coulomb potential includes the reaction potential---representing the response of the solvent by polarization and rearrangement of free ions---and any effects from the immersed surface.
We can also interpret it as the potential generated by the boundary $\Gamma$ of the molecular region $\Omega$. Taking the first expression of Equation \eqref{eq:green_identity} and subtracting out the Coulombic effect yields
\begin{equation} \label{eq:phi_reac_bem}
\phi_{\text{reac},\mathbf{r}_k} = -K_{L}^{\mathbf{r}_k}(\phi_{1,\Gamma_1}) + V_{L}^{\mathbf{r}_k} \left(\frac{\partial}{\partial \mathbf{n}}\phi_{1,\Gamma_1} \right) 
\end{equation}

Equation \eqref{eq:solv_energy} requires evaluating $\phi_{\text{reac}}$ for each point-charge location $\mathbf{r}_k$. We obtain this by discretizing Equation \eqref{eq:phi_reac_bem} and using the solution of the linear system in Equation \eqref{eq:matrix_phi} or Equation \eqref{eq:matrix_dphi} as inputs.

\medskip
\paragraph*{Surface free energy}

Chan and co-workers \cite{ChanMitchell1983,CarnieChan1993} derived the free energy for a surface with a set charge or potential. They describe the free energy on a surface as

\begin{align} \label{eq:energy_surf}
F &= \frac{1}{2} \int_{\Gamma} G_c \sigma_0^2 d\Gamma \quad \text{ for set charge, and} \nonumber \\
F &= -\frac{1}{2} \int_{\Gamma} G_p \phi_0^2 d\Gamma \quad \text{ for set potential,}
\end{align} 

\noindent where $\phi_0$ and $\sigma_0$ are the prescribed potential and surface charge, respectively. The potential is given by $\phi(\sigma, R, \mathbf{x}) = G_c(R, \mathbf{x}) \sigma$ for the first expression and the surface charge by $\sigma(\phi, R, \mathbf{x}) = G_p(R, \mathbf{x}) \phi$ for the second one. This is valid because we are using a linearized Poisson-Boltzmann model.

Using constant values of $\phi$ and $\frac{\partial \phi}{\partial \mathbf{n}}$ per panel, the discretized version of Equation \eqref{eq:energy_surf} takes the form

\begin{align} \label{eq:energy_surf_disc}
F &= \frac{1}{2} \sum_{j=1}^{N_p} \phi(\mathbf{r}_j) \sigma_{0j} A_j \text{, and } \nonumber \\
F &= -\frac{1}{2} \sum_{j=1}^{N_p} \phi_{0j} \sigma(\mathbf{r}_j) A_j. 
\end{align}

\noindent where $A_j$ is the area of panel $j$, and $\sigma = \epsilon \frac{\partial \phi}{\partial \mathbf{n}}$. To obtain the surface free energy, we can plug in the solution of the system in Equation \eqref{eq:matrix_phi} or \eqref{eq:matrix_dphi} to Equation \eqref{eq:energy_surf_disc}. 

\medskip
\paragraph*{Interaction free energy}
When there are two or more bodies in the solvent, they will interact electrostatically. In order to compute the energy of interaction, we need to take the difference between the total energy of the interacting system and the total energy of each isolated component, where the total free energy is given by
\begin{equation} \label{eq:total_energy}
F_{\text{total}} = F_{\text{Coulomb}} + F_{\text{surface}} + F_{\text{solv}}.
\end{equation}

\noindent The interaction free energy is
\begin{equation} \label{eq:interaction_energy}
F_{\text{interaction}} = F_{\text{total}}^{\text{assembly}} - \sum_{i=1}^{N_c} F_{\text{total}}^{\text{comp}_i},
\end{equation}

\noindent where $N_c$ is the number of components in the system and $F_{\text{total}}^{\text{comp}_i}$ is calculated over the isolated component $i$.


\section{Analytical solution} \label{sec:analytical_solution}

It is possible to derive a closed-form expression for the free energy of interaction between a spherical molecule with a centered charge and a spherical surface with imposed potential or charge, like the one sketched in Figure \ref{fig:twosphere_an}.  There are such analytical expressions for interacting charged surfaces \cite{CarnieChanGunning1994}, and interacting spherical molecules with multiple point charges inside \cite{LotanHead-Gordon2006}, but not for a situation where surfaces and molecules interact. Having such an analytical solution is of great utility in the development of a computational model for protein-surface interaction, because it will allow for proper code verification.

\subsection{Expansion in Legendre polynomials} \label{sec:expansion_analytical}

\begin{figure}
   \centering
   \includegraphics[width=0.45\textwidth]{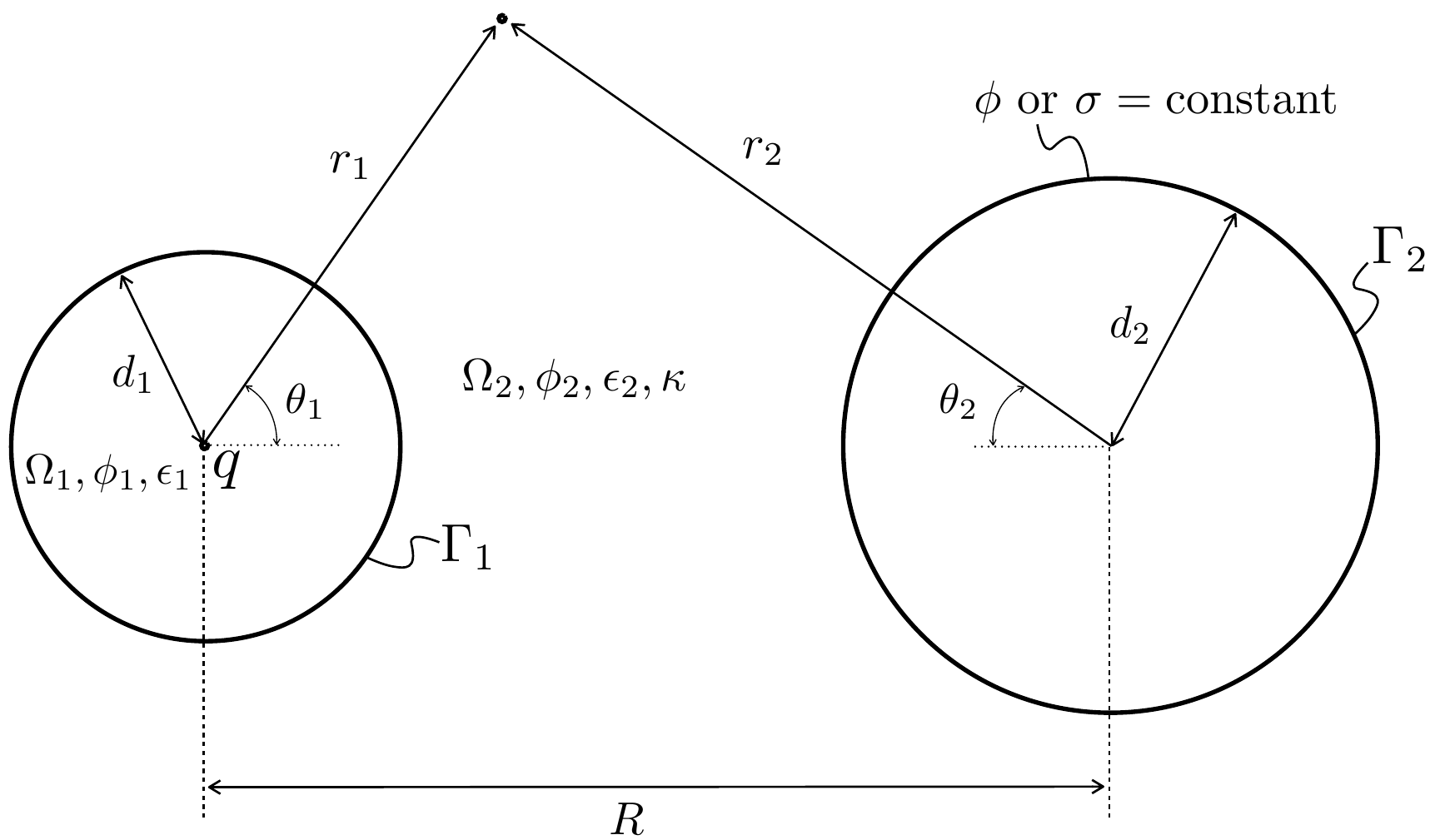} 
   \caption{Sketch of system solved with Legendre polynomials expansions.}
   \label{fig:twosphere_an}
\end{figure}

The system of partial differential equations from Equation \eqref{eq:pde}  models the electrostatic potential field in the setting of Figure \ref{fig:twosphere_an}. Following Carnie and co-workers \cite{CarnieChanGunning1994}, the axial symmetry lets us formulate the solution of Equation \eqref{eq:pde} as an expansion in Legendre polynomials:

\begin{align} \label{eq:derivation1}
\phi_1 = \sum_{n=0}^{\infty} c_n r_1^n P_n(\cos \theta_1) & + \frac{q}{4\pi\epsilon_1 r_1} \quad \text{on $\Omega_1$,} \nonumber \\
\phi_2 = \sum_{n=0}^{\infty} a_n k_n(\kappa r_1) P_n (& \cos \theta_1) \nonumber \\
+ \sum_{n=0}^{\infty} b_n k_n & (\kappa r_2) P_n(\cos \theta_2) \quad \text{ on $\Omega_2$,}
\end{align}

\noindent being $P_n$ the $n^{\text{th}}$-degree Legendre polynomial and $k_n$ the modified spherical Bessel function of the second kind.

We make use of the following addition formula \cite{MarceljaMitchellNinhamSculley1977},
\begin{equation} \label{eq:addition_formula}
k_n(\kappa r_2) P_n(\cos \theta_2) = \sum_{m=0}^{\infty}(2m+1) B_{nm} i_m(\kappa r_1) P_m(\cos \theta_1),
\end{equation}
\noindent to reformulate the expression for $\phi_2$ in Equation \eqref{eq:derivation1} as 
\begin{align} \label{eq:derivation2}
\phi_2 =& \sum_{n=0}^{\infty} a_n k_n(\kappa r_1) P_n(\cos \theta_1) \nonumber \\
& + \sum_{n=0}^{\infty} b_n \sum_{m=0}^{\infty}(2m+1) B_{nm} i_m(\kappa r_1) P_m(\cos \theta_1) \nonumber \\ 
\phi_2 =& \sum_{n=0}^{\infty} b_n k_n(\kappa r_2) P_n(\cos \theta_2) \nonumber \\
& + \sum_{n=0}^{\infty} a_n \sum_{m=0}^{\infty}(2m+1) B_{nm} i_m(\kappa r_2) P_m(\cos \theta_2).
\end{align}

Here, $i_m$ is the modified spherical Bessel function of the first kind; $B_{nm}$ is defined by 

\begin{equation} \label{eq:Bnm}
B_{nm} = \sum_{\nu=0}^{\infty} A_{nm}^{\nu} k_{n+m-2\nu}(\kappa R),
\end{equation}

\noindent where $R$ is the center-to-center distance; and $A_{nm}^{\nu}$ is given by the following expression, with ${\bf \Gamma}_x={\bf \Gamma}(x)$ (in this context only) representing the gamma function:

\begin{equation} \label{eq:Anm}
A_{nm}^{\nu} = \frac{{\bf \Gamma}_{n-\nu+0.5}{\bf \Gamma}_{m-\nu+0.5}{\bf \Gamma}_{\nu+0.5}(n+m-\nu)!(n+m-2\nu+0.5)}{\pi {\bf \Gamma}_{m+n-\nu+1.5}(n-\nu)!(m-\nu)!\nu!}.
\end{equation}

Legendre polynomials are orthogonal to each other, and $\frac{q}{4\pi\epsilon_1 r_1}$ is independent of $\theta$. Thus, taking the inner product of the expressions in Equations \eqref{eq:derivation1} and  \eqref{eq:derivation2} with $P_j(\cos \theta_i)$, where $i=1$ or $2$, yields
\begin{equation} \label{eq:derivation3}
\phi_1\delta_{0j} = c_j r_1^j + \frac{q}{4\pi\epsilon_1 r_1} \delta_{0j}  
\end{equation}
\noindent for the first expression of Equation \eqref{eq:derivation1}, and
\begin{align} \label{eq:derivation3.5}
\phi_2\delta_{0j} = &a_j k_j(\kappa r_1) + \sum_{n=0}^{\infty} b_n(2j+1)B_{nj} i_j(\kappa r_1),  \nonumber \\
\phi_2\delta_{0j} = &b_j k_j(\kappa r_2) + \sum_{n=0}^{\infty} a_n(2j+1)B_{nj} i_j(\kappa r_2)  
\end{align}
\noindent for Equation \eqref{eq:derivation2}.

Applying the interface conditions for $\Gamma_1$ on Equation \eqref{eq:derivation3} and the first expression of Equation \eqref{eq:derivation3.5}, produces
\begin{align}\label{eq:derivation4}
\sum_{n=0}^{\infty} a_n \left( \kappa k_{n}'(\kappa d_1) - \frac{\epsilon_1}{\epsilon_2} \frac{n}{d_1} k_n(\kappa d_1) \right) \delta_{nj} +& \nonumber \\ 
b_n (2j+1)B_{nj} \left( \kappa i_{j}'(\kappa d_1) - \frac{\epsilon_1}{\epsilon_2} \frac{j}{d_1} i_j(\kappa d_1)  \right) & = \nonumber \\
-\frac{\epsilon_1}{\epsilon_2} \frac{q}{4\pi\epsilon_1 d_1^2} \delta_{0j}&(j+1),
\end{align}

\noindent where $d_1$ is the radius of surface $1$.

\subsubsection*{Constant potential $\phi$ on $\Gamma_2$.}
The application of the boundary condition on $\Gamma_2$, $\phi(\Gamma_2) = \phi_0$, where $\phi_0$ is independent on $\theta_2$, gives

\begin{equation} \label{eq:derivation5_phi}
\sum_{n=0}^{\infty} a_n(2j+1)B_{nj}i_j(\kappa d_2) + b_nk_n(\kappa d_2) \delta_{nj} = \phi_0 \delta_{0j}.
\end{equation}

\noindent Combining Equations \eqref{eq:derivation4} and  \eqref{eq:derivation5_phi} yields the following system of equations for the coefficients $a_n$ and $b_n$
\begin{align} \label{eq:system_phi}
\mathbf{I} \mathbf{A} + \mathbf{L} \mathbf{B} &= -\frac{\epsilon_1}{\epsilon_2} \frac{q}{4\pi\epsilon_1 d_1^2} \mathbf{e} \nonumber \\
\mathbf{M} \mathbf{A} + \mathbf{I} \mathbf{B} &= \phi_0 \mathbf{e}
\end{align}

\noindent where
\begin{align} \label{eq:phi_terms}
I_{jn} &= \delta_{jn} \nonumber \\
e_j &= \delta_{0j} \nonumber \\
A_n &= a_n \left(\kappa k_n'(\kappa d_1) - \frac{\epsilon_1}{\epsilon_2} \frac{n}{d_1} k_n(\kappa d_1) \right) \nonumber \\
B_n &= b_n k_n(\kappa d_2) \nonumber \\
L_{jn} &= (2j+1)B_{nj}\left( \kappa \frac{i_j'(\kappa d_1)}{k_n(\kappa d_2)} - \frac{\epsilon_1}{\epsilon_2} \frac{j}{d_1} \frac{i_j(\kappa d_1)}{k_n(\kappa d_2)} \right) \nonumber \\
M_{jn} &= (2j+1)B_{nj} i_j(\kappa d_2) \frac{1}{\left(\kappa k_n'(\kappa d_1) - \frac{\epsilon_1}{\epsilon_2} \frac{n}{d_1} k_n(\kappa d_1) \right)}. 
\end{align}

\subsubsection*{Constant surface charge $\sigma$ on $\Gamma_2$.}
In this case, the application of the boundary condition on $\Gamma_2$, $\sigma(\Gamma_2) = -\epsilon_2 \frac{\partial \phi}{\partial \mathbf{n}} \Large|_{\Gamma_2} = \sigma_0$, where $\sigma_0$ is independent on $\theta_2$, gives

\begin{equation} \label{eq:derivation5_dphi}
\sum_{n=0}^{\infty} a_n(2j+1)B_{nj}\kappa i_j'(\kappa d_2) + b_n \kappa k_n'(\kappa d_2) \delta_{nj} = -\frac{\sigma_0}{\epsilon_2} \delta_{0j}
\end{equation}

\noindent Combining Equations \eqref{eq:derivation4} and  \eqref{eq:derivation5_phi} produces a system of equations for the coefficients $a_n$ and $b_n$
\begin{align} \label{eq:system_dphi}
\mathbf{I} \mathbf{A} + \mathbf{L} \mathbf{B} &= -\frac{\epsilon_1}{\epsilon_2} \frac{q}{4\pi\epsilon_1 d_1^2} \mathbf{e} \nonumber \\
\mathbf{M} \mathbf{A} + \mathbf{I} \mathbf{B} &= -\frac{\sigma_0}{\epsilon_2} \mathbf{e}
\end{align}

\noindent where
\begin{align} \label{eq:dphi_terms}
I_{jn} &= \delta_{jn} \nonumber \\
e_j &= \delta_{0j} \nonumber \\
A_n &= a_n \left(\kappa k_n'(\kappa d_1) - \frac{\epsilon_1}{\epsilon_2} \frac{n}{d_1} k_n(\kappa d_1) \right) \nonumber \\
B_n &= b_n \kappa k_n'(\kappa d_2) \nonumber \\
L_{jn} &= (2j+1)B_{nj}\left( \frac{i_j'(\kappa d_1)}{k_n'(\kappa d_2)} - \frac{\epsilon_1}{\epsilon_2} \frac{j}{d_1} \frac{i_j(\kappa d_1)}{\kappa k_n'(\kappa d_2)} \right) \nonumber \\
M_{jn} &= (2j+1)B_{nj} \kappa i_j'(\kappa d_2) \frac{1}{\left(\kappa k_n'(\kappa d_1) - \frac{\epsilon_1}{\epsilon_2} \frac{n}{d_1} k_n(\kappa d_1) \right)}. 
\end{align}
 
\subsection{Energy calculation} \label{energy_analytical}

 \medskip
 \paragraph*{Solvation free energy of the molecule}
According to Equation \eqref{eq:solv_energy}, the solvation free energy of a molecule with a centered charge is given by
\begin{equation} \label{eq:energy_phi}
F_{\text{solv}} = \frac{1}{2} q \phi_{\text{reac}}(r_1=0),
\end{equation} 
 
 \noindent and using Equation \eqref{eq:derivation1}, the reaction potential from Equation \eqref{eq:phi_reac_bem} is:
 \begin{equation} \label{eq:phi_reac_an}
 \phi_{\text{reac}} = \phi - \frac{q}{4\pi\epsilon_1 r} = \sum_{n=0}^{\infty} c_n r^n P_n(\cos \theta_1).
 \end{equation}
 
 Applying the boundary conditions at $\Gamma_1$ on Equation  \eqref{eq:derivation3}, we can rewrite $c_j$ in terms of the already computed $a_j$ and $b_j$:
 \begin{align}
 c_j = \frac{1}{d_1^j} & \Big(a_j k_j(\kappa d_1) + \nonumber \\
&  \sum_{m=0}^{\infty} b_m(2j+1)B_{mj} i_j(\kappa d_1) - \frac{q}{4\pi\epsilon_1 d_1} \delta_{0j} \Big)
 \end{align} 
 
Because the charge is located at $r=0$, only the $n=0$ terms of Equation \eqref{eq:phi_reac_an} will survive, and the potential at this location is:
 
 \begin{align} \label{eq:phi_reac_an2}
 \phi_{\text{reac}} (r_1=0) = & a_0 k_0(\kappa d_1) + \nonumber \\
 &\sum_{m=0}^{\infty} b_m B_{m0}i_0(\kappa d_1) - \frac{q}{4\pi\epsilon_1 d_1}
 \end{align}
 
 The result from Equation \eqref{eq:phi_reac_an2} in Equation \eqref{eq:energy_phi} yields the solvation free energy. 
 
For the isolated molecule, $R \to \infty$ makes $B_{nm} \to 0$, which nullifies the sum in Equation \eqref{eq:phi_reac_an2} and $a_0$ for $R \to \infty$, from the system in Equation \eqref{eq:system_phi}, is 

\begin{equation} \label{eq:a0_inf}
a_0^{\infty} = -\frac{q}{d_1^2}\frac{\epsilon_1}{\epsilon_2} \frac{1}{4\pi\kappa k_0'(\kappa d_1) \epsilon_1}
\end{equation}

\medskip
\paragraph*{Surface  free energy with set potential $\phi_0$}
We can expand $G_p$ from Equation \eqref{eq:energy_surf} in Legendre polynomials as
\begin{align} \label{eq:G_p}
G_p = &-\frac{\epsilon_2 \kappa}{\phi_0}  \Bigg[ \sum_{n=0}^{\infty} b_n k_n'(\kappa d_2) P_n(\cos \theta_2) \nonumber \\ 
& + \sum_{n=0}^{\infty} a_n \sum_{m=0}^{\infty} (2m+1) B_{nm} i_m'(\kappa d_2) P_m(\cos \theta_2) \Bigg].
\end{align}

 \noindent Applying Equation \eqref{eq:G_p} in Equation \eqref{eq:energy_surf} gives

\begin{equation} \label{G_p_int}
F = 2\pi \kappa \phi_0 d_2^2 \epsilon_2 \left[ b_0 k_0'(\kappa d_2) + \sum_{n=0}^{\infty} a_n B_{n0} i_0'(\kappa d_2) \right]
\end{equation}

 \noindent If the surface is isolated, $R \to \infty$ makes $B_{n0} \to 0$, and the free energy in this case is 
\begin{equation} \label{energy_isolated_phi}
F = 2\pi \kappa \phi_0 d_2^2 b_0^{\infty} k_0'(\kappa d_2) \epsilon_2
\end{equation}
 
 \noindent where $b_0^{\infty}$ is taken from the system in  \eqref{eq:system_phi} considering $B_{nm} \to 0$, which results in
 \begin{equation} \label{b_inf_phi}
 b_0^{\infty} = \frac{\phi_0}{k_0(\kappa d_2)}.
 \end{equation}
 
 \medskip
 \paragraph*{Surface  free energy with set charge $\sigma_0$}
We can expand $G_c$ from Equation \eqref{eq:energy_surf} in Legendre polynomials as
\begin{align} \label{eq:G_c}
G_c = & \frac{1}{\sigma_0} \Bigg[ \sum_{n=0}^{\infty} b_n k_n(\kappa d_2) P_n(\cos \theta_2) + \nonumber \\ 
&\sum_{n=0}^{\infty} a_n \sum_{m=0}^{\infty} (2m+1) B_{nm} i_m(\kappa d_2) P_m(\cos \theta_2) \Bigg]
\end{align}

 \noindent Applying Equation \eqref{eq:G_c} into Equation \eqref{eq:energy_surf} gives

\begin{equation} \label{G_c_int}
F = 2\pi \sigma_0 d_2^2 \left[ b_0 k_0(\kappa d_2) + \sum_{n=0}^{\infty} a_n B_{n0} i_0(\kappa d_2) \right]
\end{equation}

 \noindent For the isolated surface, $R \to \infty$ and $B_{n0} \to 0$, and the free energy is 
\begin{equation} \label{energy_isolated_dphi}
F = 2\pi \sigma_0 d_2^2 b_0^{\infty} k_0(\kappa d_2) 
\end{equation}
 
 \noindent where $b_0^{\infty}$ is calculated from the system in  \eqref{eq:system_dphi} considering $B_{nm} \to 0$, which results in
 
 \begin{equation} \label{b_inf_dphi}
 b_0^{\infty} = -\frac{\sigma_0}{\epsilon_2 \kappa k_0'(\kappa d_2)}.
 \end{equation}

\section{Results of the grid-convergence study} \label{sec:results}

To obtain the following results, we extended the \pygbe code to consider surfaces with prescribed charge or potential. For all runs, we used a workstation with Intel Xeon X5650 \cpu s  and one \nvidia Tesla C2075 \gpu\ card (2011 Fermi). We used the free \msms software \cite{SannerOlsonSpehner1995} to generate meshes, and \texttt{pdb2pqr} \cite{Dolinsky04} with an \amber force field to determine the charges and van der Waals radii. 

\subsection{Verification against analytical solution} \label{sec:verification}

\begin{figure}[h] 
   \centering
   \includegraphics[width=0.45\textwidth]{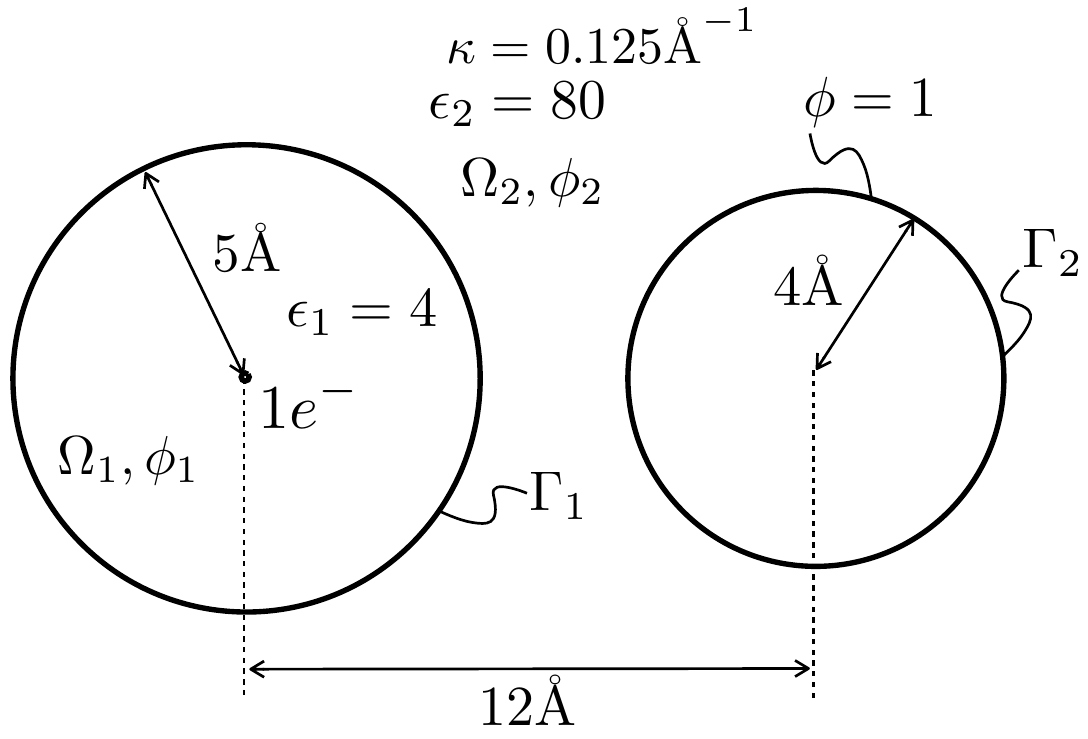} 
   \caption{Sketch of system used in the convergence study of Figure \ref{fig:error_sphere}.}
   \label{fig:twosphere_num}
\end{figure}

Using the analytical solution detailed in Section \ref{sec:analytical_solution}, we carried out a grid-convergence study of \pygbe extended to treat interacting surfaces with biomolecules. The setup consists of a spherical molecule with a $5$\AA~radius and a centered charge of $1e^-$, interacting with a spherical surface of $4$\AA~radius and an imposed potential of $\phi=1$. The center-to-center distance between the spheres is $12$\AA, and they are dissolved in water with salt at 145mM, which gives a Debye length of 8 ($\kappa = 0.125$), and permittivity $\epsilon_\text{sol} = 80$. The permittivity inside the spherical protein is $\epsilon_\text{mol} = 4$. Figure \ref{fig:twosphere_num} shows a sketch of this system.

\begin{figure}[htbp] 
   \centering
   \includegraphics[width=0.4\textwidth]{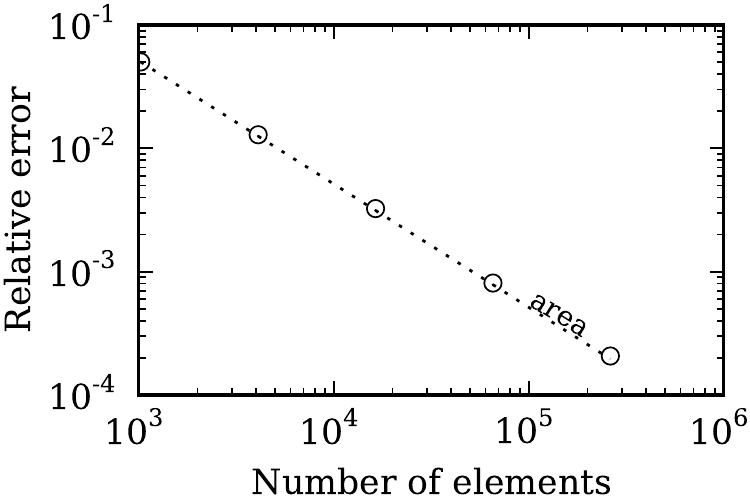} 
   \caption{Grid-convergence study for the interaction free energy between a spherical molecule with a centered charge and a sphere with potential $\phi=1$. Data sets, figure files plus running/plotting scripts are available under \ccby \cite{CooperBarba2015-share1348841}.}
   \label{fig:error_sphere}
\end{figure}

Figure \ref{fig:error_sphere} presents the results of the grid-convergence analysis, where the error is the relative difference in interaction free energy between the analytical result from Section \ref{sec:analytical_solution} and the numerical solution computed with \pygbe. The observed order of convergence of the three finest meshes was 1.007. Table \ref{table:params1} presents the numerical parameters used in this case. Recall from section \ref{sec:methods} that we calculate the boundary-element integrals differently for close-by and far-away elements, and use a semi-analytical method for the element that contains the collocation point. The fine Gauss quadrature rule is used for elements closer than $2L$ from the collocation point, where $L=\sqrt{2\cdot \text{Area}}$. For the treecode,  $N_{\text{crit}}$ is the maximum number of boundary elements per box, $P$ is the Taylor expansion truncation parameter and $\theta$ is the multipole-acceptance criterion. The final numerical parameter is the exit tolerance of the \textsc{gmres} solver.

\begin{table}[h]
   \caption{\label{table:params1}Numerical parameters used in the code-verification runs with the analytical solution. } 
    \begin{tabular}{c c c c c c c}
	\hline
	\multicolumn{3}{l} {\# Gauss points:} & \multicolumn{3}{l}{Treecode:} & \gmres:\\
	\footnotesize{in-element} & \footnotesize{close-by} & \footnotesize{far-away} & $N_{\text{crit}}$ & $P$ &  $\theta$  & tol.\\
	\hline
	9 per side & 37 & 3  &  300 & 15 & 0.5  & $10^{-9}$\\	
	\hline
    \end{tabular}
\end{table}

As seen in Figure \ref{fig:error_sphere}, the error decays with the average area of the boundary elements ($\frac{1}{N}$), which is the expected behavior and consistent with our previous verification exercises \cite{CooperBarba-share154331}. This proves that the extension of \pygbe to treat charged surfaces is solving the mathematical model correctly.

\begin{figure}
   \centering
   \includegraphics[width=0.4\textwidth]{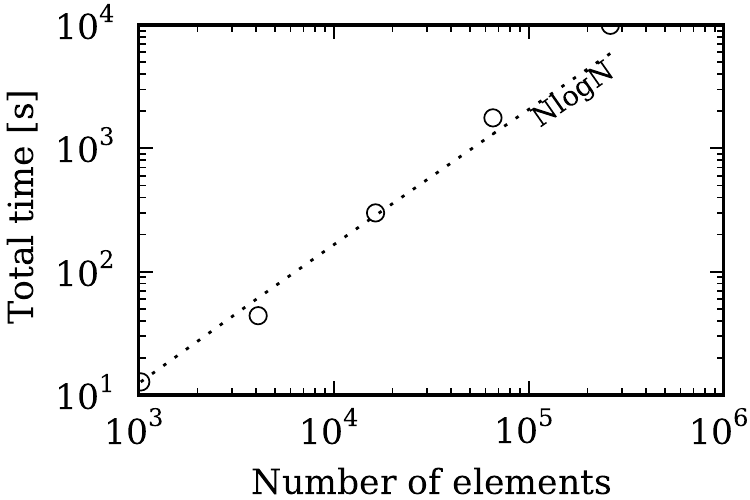} 
   \caption{Total runtime for obtaining interaction free energy (requiring three separate runs: one for each surface in isolation and one for both together), for the two-sphere system of \ref{sec:verification}.}
   \label{fig:time}
\end{figure}

\begin{figure}
   \centering
   \includegraphics[width=0.4\textwidth]{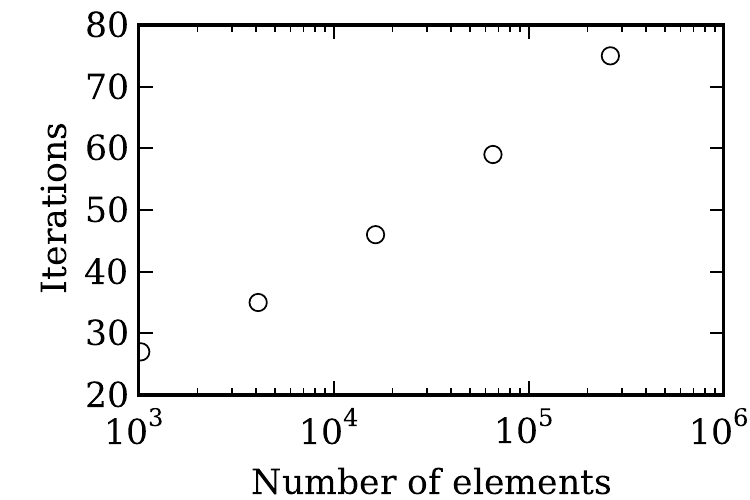} 
   \caption{Number of iterations to converge for the two-sphere system of \ref{sec:verification}. Iteration count increases with problem size due to the conditioning properties of the boundary integral formulation in our model.}
   \label{fig:iterations}
\end{figure}

Obtaining the interaction free energy involves three separate calculations: one with both bodies (molecule and interacting surface with set potential) and one for each isolated body. Figure \ref{fig:time} shows the total time to solution for each mesh, including all three cases that need to be computed. The most time-consuming part of the algorithm is the matrix-vector product within the Krylov solver, which scales as $O(N \log N)$ thanks to the treecode acceleration. However, the total time-to-solution scales slightly worse than $O(N \log N)$ because the condition number of the system increases with $N$, and we need more iterations of the Krylov solver to converge to the desired tolerance, as shown in Figure \ref{fig:iterations}.

\subsection{Protein \gb} \label{sec:PGB}

We computed the electrostatic field of protein \gb interacting with a 100\AA$\times$100\AA$\times$10\AA\ block with surface charge density $0.05$C/m$^2$. 
The protein was centered with respect to a  100\AA$\times$100\AA\ face, a distance 2\AA\ above it. Since we did not consider any Stern layers or solvent-filled cavities, these tests contain only two surfaces: the protein's \ses and the charged surface.
We also computed the electrostatic field generated by protein \gb and the surface by themselves.

\begin{figure}
   \centering
   \includegraphics[width=0.35\textwidth]{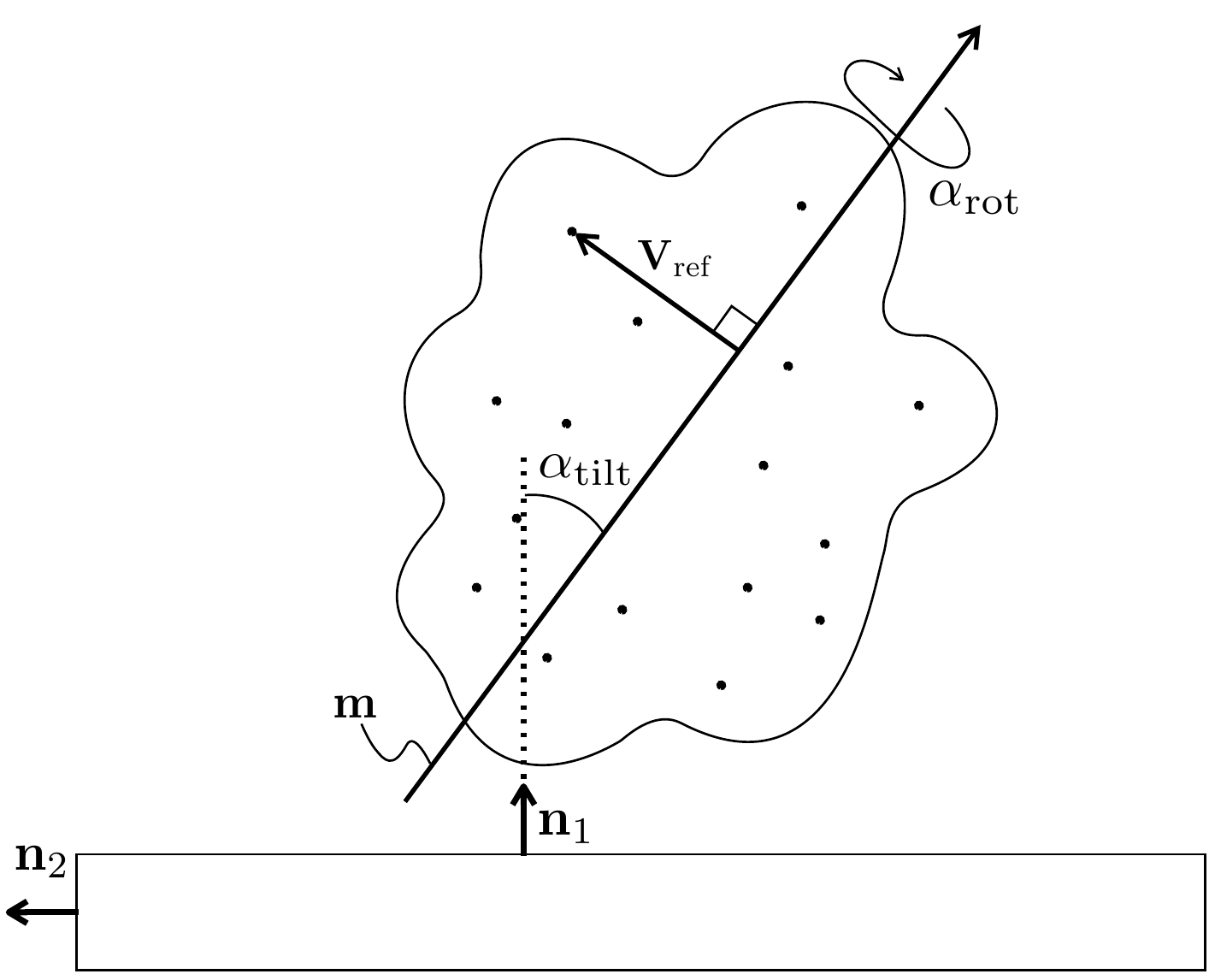} 
   \caption{Orientation of a protein near a charged surface. $\mathbf{m}$ is the dipole moment vector, $\mathbf{V}_\text{ref}$ the vector between $\mathbf{m}$ and the atom that is the furthest, and $\mathbf{n}_1$ and $\mathbf{n}_2$ are normal to their corresponding surfaces. $\alpha_\text{tilt}$ is the angle between $\mathbf{n}_1$ and $\mathbf{m}$, and $\alpha_\text{rot}$ the angle between $\mathbf{V}_\text{ref}$ and $\mathbf{n}_2$ when $\mathbf{m}$ and $\mathbf{n}_1$ are aligned.}
   \label{fig:protein_surface}
\end{figure}

The angle between the dipole moment of the protein and the vector normal to the surface was $\alpha_\text{tilt}=10^\circ$. 
The dipole-moment vector placed at the center of mass of the protein generates an axis, and we used the line of shortest distance between the outermost atom and this axis as a reference vector $\mathbf{V}_{\text{ref}}$. 
The rotation angle $\alpha_{\text{rot}}$ is the angle between the normal vector to a 100\AA$\times$10\AA\ side face of the block and $\mathbf{V}_{\text{ref}}$ when $\alpha_\text{tilt}=0$, and is equal to $200^\circ$ in these tests. Figure \ref{fig:protein_surface} is a sketch of this arrangement.

\begin{figure}
   \centering
   \includegraphics[width=0.45\textwidth]{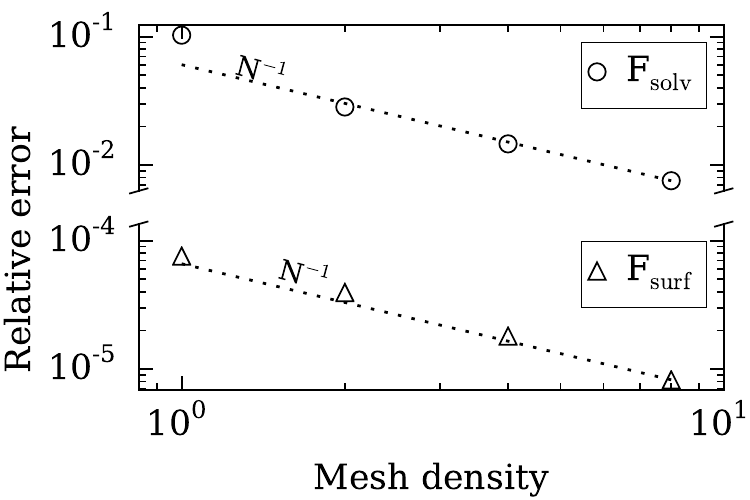} 
   \caption{Grid-convergence study of the solvation energy for an isolated protein \gb mutant, and the surface energy of an isolated surface with charge density of 0.05C/m$^2$.}
   \label{fig:convergence_1PGB_isolated}
\end{figure}

\begin{figure}[h] 
   \centering
   \includegraphics[width=0.45\textwidth]{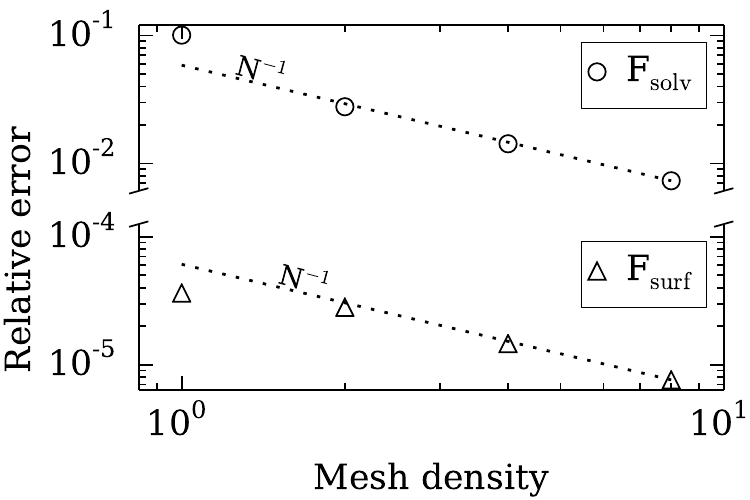} 
   \caption{Grid-convergence study of the solvation and surface energy for protein \gb mutant, interacting with a surface with a charge density of 0.05C/m$^2$. Data sets, figure files and running/plotting scripts available under \ccby \cite{CooperBarba2015-share1348803}}.
   \label{fig:convergence_1PGB_sensor}
\end{figure}

In this test, the solvent has no salt, i.e., $\kappa=0$, and its relative permittivity was 80. The region inside the protein had a relative permittivity of 4.
We computed the solvation and surface energy using meshes with 1, 2, 4 and 8 elements per square Angstrom with the parameters detailed in Table \ref{table:params2}. 
Using Richardson extrapolation and the result of the three finest meshes we calculated an approximate exact solution, shown in Table \ref{table:extraPGB}, which we consider as a reference to calculate estimated errors. 
The errors plotted in Figures \ref{fig:convergence_1PGB_isolated} and \ref{fig:convergence_1PGB_sensor} are the relative differences between the energy obtained with Richardson extrapolation and the results computed with each mesh.  
They decay as $1/N$ in both the solvation and surface energies for the finest three meshes, indicating that the calculations are in the asymptotic region and the geometry is well resolved in these cases.
The observed order of convergence is 0.95 for the solvation energy and 1.12 for the surface energy in the isolated cases, and 0.96 for the solvation energy and 0.94 for the surface energy when the protein and surface were interacting. 
Using the extrapolated values from Table \ref{table:extraPGB}, we obtain an interaction free energy of $-7.6$ [kcal/mol].
For details on the Richardson-extrapolation method for performing grid-convergence analysis, see our previous work \cite{CooperBardhanBarba2013}.

\begin{table}[h]
   \caption{\label{table:params2}Numerical parameters used in the convergence runs with protein \gb. } 
    \begin{tabular}{c c c c c c c}
	\hline
	\multicolumn{3}{l} {\# Gauss points:} & \multicolumn{3}{l}{Treecode:} & \gmres:\\
	\footnotesize{in-element} & \footnotesize{close-by} & \footnotesize{far-away} & $N_{\text{crit}}$ & $P$ &  $\theta$  & tol.\\
	\hline
	9 per side & 19 & 7  &  500 & 15 & 0.5  & $10^{-8}$\\	
	\hline
    \end{tabular}
\end{table}

\begin{table}[h]
   \caption{\label{table:extraPGB}Extrapolated values of energy for protein \gb.} 
    \begin{tabular}{c c c}
	\hline
	& \multicolumn{2}{c} {Energy [kcal/mol]} \\
	& Solvation & Surface \\
	\hline
    Isolated    & $-218.26$ & $317.41$ \\
	Interacting & $-222.43$ & $313.98$ \\	
	\hline
    \end{tabular}
\end{table}

\subsection{Reproducibility and data management}
To facilitate the replication of our work, we consistently release code and data associated with every publication. In that context, \pygbe was released under an MIT open-source license with our previous publication \cite{CooperBardhanBarba2013}, and continues to be available via its version-control repository. 
Supplementing this paper, we prepared \emph{``reproducibility packages''} containing running and post-processing scripts in Python to generate Figures \ref{fig:error_sphere} and \ref{fig:convergence_1PGB_sensor}. The packages invoke \pygbe with the parameters and meshes reported here, and then produce the plots, all with a single command.
The reproducibility packages are hosted on \textbf{figshare}, and are referenced in the respective captions.

\section{Discussion} \label{sec:discussion}

In order to study the interaction of proteins and charged surfaces, we extended \pygbe to account for surfaces with prescribed charge or potential. Unfortunately, there was no analytical solution available in the literature to compare and verify \pygbe's extension. 
Section \ref{sec:analytical_solution} derives a closed-form expression for a spherical molecule with a centered charge interacting with a spherical nanosurface with imposed charge or potential.
We used this new analytical solution to conduct a grid-convergence study of the interaction energy (Figure \ref{fig:error_sphere}). The error decays with the area, which is the expected behavior for a boundary element method with constant elements \cite{CooperBardhanBarba2013, CooperBarba-share154331}. 
Discretization error is very small for a spherical geometry. To make sure that the errors due to integration, the treecode approximation and the \textsc{gmres} solver were even smaller, we chose all the numerical parameters for high accuracy. This allows us to observe the convergence with respect to the discretization only and extract the order. With more realistic molecular geometries, however, discretization errors will be larger and the requirements can be relaxed in the other numerical parameters of \pygbe, resulting in lower runtimes.


The results in Figures \ref{fig:convergence_1PGB_isolated} and \ref{fig:convergence_1PGB_sensor} show the applicability of this approach in more realistic situations. The setting in Figure \ref{fig:protein_surface} can model a nanostructure coated with a self-assembled monolayer (\sam) interacting with a protein that will adsorb on that surface. Our application of interest in developing this model is the field of biosensors, where it can assist design through studies of electrostatic adsorption affecting protein orientation near the biosensor. With antibody-based biosensors, orientation determines the accessibility of reaction sites and is critical for sensitivity. In a companion publication \cite{CooperBarba2015b}, we present the first studies of protein orientation near charged surfaces using our modeling framework.

Figures \ref{fig:convergence_1PGB_isolated} and \ref{fig:convergence_1PGB_sensor} show the expected 1/N convergence with a simple protein. Just like in the case of the sphere, the simple geometry of the charged surface forced us to use very fine parameters in order to extract the order of convergence. If we needed to run this computation many times---for example, to sample different protein orientations---we might relax these parameters to obtain shorter computation times. 

The extrapolated values in Table \ref{table:extraPGB} are useful to find more relaxed parameters that still give acceptable results. For example, using a mesh density of 2 elements per square Angstrom on the charged surface and 4 elements per square Angstrom on the protein, and the parameters detailed in Table \ref{table:params3}, we get the results in Table \ref{table:relaxPGB}. These results are less than 2\% away from those in Table \ref{table:extraPGB}, and each run takes less than one minute. Moreover, using the energy values from Table \ref{table:relaxPGB}, the interaction free energy is $-7.61$ [kcal/mol], which is very close to the extrapolated case ($-7.6$ [kcal/mol]).

\begin{table}[h]
   \caption{\label{table:params3}Numerical parameters for relaxed runs with protein \gb. } 
    \begin{tabular}{c c c c c c c}
    \hline
    \multicolumn{3}{l} {\# Gauss points:} & \multicolumn{3}{l}{Treecode:} & \gmres:\\
    \footnotesize{in-element} & \footnotesize{close-by} & \footnotesize{far-away} & $N_{\text{crit}}$ & $P$ &  $\theta$  & tol.\\
    \hline
    9 per side & 19 & 1  &  300 & 4 & 0.5  & $10^{-5}$\\
    \hline
    \end{tabular}
\end{table}

\begin{table}[h]
   \caption{\label{table:relaxPGB}Values of energy for protein \gb using the parameters in Table \ref{table:params3}, and a mesh density of 4 elements per square angstrom in the protein and 2 elements per square angstrom on the charged surface}
    \begin{tabular}{c c c}
    \hline
    & \multicolumn{2}{c} {Energy [kcal/mol]} \\
    & Solvation & Surface \\
    \hline
    Isolated    & $-221.56$ & $315.33$ \\
    Interacting & $-225.81$ & $311.96$ \\
    \hline
    \end{tabular}
\end{table}

\section{Conclusion}

In this work, we used an implicit-solvent model to study protein-surface interaction. We present for the first time and apply an extension of our open-source \pygbe code to account for the presence of surfaces with imposed potential or charge. The new feature of the code was verified against an analytical solution, which we derived for that purpose. 

To demonstrate the power of this approach in a more realistic setting, we performed tests of protein G B1 D4$^\prime$ near a brick-shaped surface with an imposed charge. The error in energy scaling with the area of boundary elements demonstrates that this extension of \pygbe is capable of resolving the mathematical model correctly. This test was motivated by the biosensing application, where a ligand molecule is adsorbed on a \sam-coated nanoparticle, which can be represented by the brick-shaped surface.

The addition of a surface with imposed charge or potential in the implicit-solvent model falls naturally in a boundary integral approach. In this case, the region enclosed by the surface is not part of the domain, then, this surface only adds one equation to the linear system, rather than two, which is the case with the molecular solvent-excluded surface.

We conclude that this implicit-solvent model can offer a valuable approach in protein-surface interaction studies. This tool can be useful for orientation studies of ligand molecules in biosensors, either to find optimal adsorption conditions of salt concentration and surface charge, or to guide the design of better ligand molecules.

\section*{Acknowledgments}
 This work was supported by ONR via grant \#N00014-11-1-0356 of the Applied Computational Analysis Program. LAB also acknowledges support from NSF CAREER award OCI-1149784 and from NVIDIA, Inc.\ via the CUDA Fellows Program. 
 We are grateful for many helpful conversations with members of the Materials and Sensors Branch of the Naval Research Laboratory, especially Dr. Jeff M. Byers and Dr. Marc Raphael.


\end{document}